\documentclass[a4paper,11pt]{article}

\usepackage{float}
\usepackage{caption}
\usepackage{subcaption}
\usepackage{lineno}
\usepackage{jcappub} 

\makeatletter
\g@addto@macro\bfseries{\boldmath}
\makeatother

\arxivnumber{2412.14607} 
\title{Testing anisotropic Hubble expansion}

\author[a]{Paula~Boubel,}
\author[a]{Matthew~Colless,}
\author[b,c]{Khaled~Said,}
\author[d]{and Lister~Staveley-Smith}

\affiliation[a]{Research School of Astronomy and Astrophysics, The Australian National University, Mount Stromlo Observatory, Canberra, ACT 2611, Australia}
\affiliation[b]{School of Mathematics and Physics, University of Queensland, Brisbane, QLD 4072, Australia}
\affiliation[c]{OzGrav:~The ARC Centre of Excellence for Gravitational Wave Discovery, Hawthorn, VIC 3122, Australia}
\affiliation[d]{International Centre for Radio Astronomy Research (ICRAR), University of Western Australia, 35 Stirling Hwy, Crawley, WA 6009, Australia}

\emailAdd{paula.boubel@anu.edu.au}

\abstract{The cosmological principle asserting the large-scale uniformity of the Universe is a testable assumption of the standard cosmological model. We explore the constraints on anisotropic expansion provided by measuring directional variation in the Hubble constant, $H_0$, derived from differential zeropoint measurements of the Tully-Fisher distance estimator. We fit various models for directional variation in $H_0$ using the Tully-Fisher dataset from the all-sky \textit{Cosmicflows-4} catalog. The best-fit dipole variation has an amplitude of 0.063\,$\pm$\,0.016\,mag in the direction ($\ell,b$) = (142\,$\pm$\,30\degree,52\,$\pm$\,10\degree). If this were due to anisotropic expansion it would imply a 3\% variation in $H_0$ (i.e.\ $\Delta H_0$\,=\,2.10\,$\pm$\,0.53\,\kmsMpc\ if $H_0$\,=\,70\,\kmsMpc) with a significance of 3.9$\sigma$. A model including this $H_0$ dipole is only weakly favored relative to a model with a constant $H_0$ and a bulk motion of the volume sampled by \textit{Cosmicflows-4} consistent with the standard $\Lambda$CDM cosmology. However, m simulations that the expected Tully-Fisher data from the WALLABY and DESI surveys should allow detection of a 1\% $H_0$ dipole anisotropy at 5.8$\sigma$ confidence and distinguish it from the typical bulk flow predicted by $\Lambda$CDM over the volume of these surveys.}

\begin{document}
\maketitle
\flushbottom

\section{Motivation}

The cosmological principle, the assumed isotropy and homogeneity of the universe on sufficiently large scales, has been put under increasing scrutiny over the past decade. Detections of an anisotropic expansion rate would create tension with this fundamental assumption of the standard model of cosmology. Hints of anisotropic expansion have been found in the quasar data \cite{Krishnan_2021, Secrest_2021, Secrest_2022} and Type~Ia supernovae data \cite{Sah_2024, Mcconville_2023, Luongo_2022, Krishnan_2022, Sorrenti_2023, Zhai_2022}. 

Ref.~\cite{Zhai_2022} found that while the amplitude of the anisotropy is not statistically unlikely, its alignment with the CMB dipole is troubling, since these supernovae compilations have already been put in the CMB frame by construction. Ref.~\cite{Sorrenti_2023} found that the direction of the $H_0$ dipole differed from that of the CMB dipole by 3$\sigma$ until a sufficiently high redshift cut was made, indicating that corrections for peculiar velocities may be extremely important in $H_0$ determinations. Ref.~\cite{Krishnan_2022} found a positive $H_0$ variation of order 1\,\kmsMpc\ in the direction of the CMB dipole, for both low- and high-redshift samples. Ref.~\cite{Luongo_2022} also found higher values of $H_0$ in the direction of the CMB dipole at 2--3$\sigma$ significance. Most recently, refs.~\cite{Sah_2024, Hu_2024, Mcconville_2023} examined the catalog of \textit{Pantheon+} Type Ia supernovae \cite{Scolnic_2022} in the CMB frame and found a dipole anisotropy $H_0$ of +2--4\,\kmsMpc\ in roughly the same direction as the CMB dipole. These studies hint at either a calibration problem or a possible misinterpretation of the CMB dipole in modern cosmology. 

A study using galaxy scaling relations \cite{Migkas_2021} found an anisotropy with a dipolar form corresponding to a 9\% spatial variation of $H_0$ in the direction ($\ell$,$b$) = (280\,$\pm$\,35\degree,$-$15\,$\pm$\,20\degree) or to a bulk flow of 900\,\kms. This direction refers to a \emph{lower} $H_0$ value compared to the rest of the sky. Using simulations, they determined that the significance of this was greater than 5$\sigma$. However, they stated that the effect of a $H_0$ dipole is inseparable from a bulk flow in their sample due to the low median redshift ($z=0.1$). Still, the large bulk flow required would be in tension with assumptions in standard $\Lambda$CDM cosmology.

Subsequently, a similar analysis was performed by \cite{Pandya_2024} while searching for systematic biases that could explain the previous result. They found no systematics large enough and their results were consistent with \cite{Migkas_2021}, finding a variation in the direction ($\ell$,$b$) = (295\,$\pm$\,71\degree,$-$30\,$\pm$\,71\degree) with a significance of 3.6$\sigma$. Both of these studies present strong evidence for an anisotropy of galaxy scaling relations, but the underlying cause could be either an anisotropic $H_0$ or a large-scale bulk flow.

Theoretical frameworks for anisotropic expansions arising from arbitrary space-time metrics beyond the standard FLRW assumption have been studied. For example, \cite{Heinesen_2021} presents a model-independent multipole expansion of cosmological luminosity distances. Using simulations of this physical framework, the parameter that appears in place of the FLRW Hubble parameter was found to be dominated by a quadrupole \cite{Cowell_2023, Macpherson_2021}. The maximum quadrupole found by \cite{Macpherson_2021} is typically 2\% but can be as high as 5\%, while \cite{Cowell_2023} found a quadrupole strength of 0.565\% on average for 100 observers. These results depend on the smoothing scale of the simulations. Using the \textit{Pantheon+} catalog \cite{Scolnic_2022} to constrain this quadrupole in the Hubble parameter, \cite{Cowell_2023} found a 1.96$\sigma$ quadrupole even with velocity corrections.

In this study, we will use the Tully-Fisher relation to investigate this recurring theme and discover whether it is possible to disentangle the effect of bulk flows from a true $H_0$ anisotropy. As with the Type~Ia supernovae studies, but in contrast to the galaxy scaling relation studies, we define the `direction' of the $H_0$ dipole to be that of its maximum value. 

The advantage of probing $H_0$ anisotropy, as opposed to the isotropic value of $H_0$, is that differential measurements are not subject to systematics in the absolute calibration of $H_0$. In fact, the presence of an $H_0$ anisotropy in the local Universe could have significant implications in the form of bias or additional sample variance for $H_0$ measurements that do not account for this possibility, since when isotropy is assumed, sky coverage is not typically considered in determinations of $H_0$. 

Although the use of the Tully-Fisher relation suffers from significant systematic errors in the determination of a $H_0$ monopole \cite{Boubel_2024b}, its ability to detect $H_0$ variations is limited only by the statistical precision of the Tully-Fisher zeropoint. At present we may not have the precision required to adequately constrain anisotropic Hubble expansions, but the bounty of new Tully-Fisher data in the next few years may make useful constraints possible in the near future. This study is a proof-of-concept demonstrating that differential Tully-Fisher zeropoint analysis will be a viable tool for detecting $H_0$ anisotropy with upcoming datasets.

In Section~\ref{sec:data} we describe the data used for this analysis; in Section~\ref{sec:method} we describe the model used to constrain the anisotropies and its integration into our Bayesian methodology; in Section~\ref{sec:results} we present the results for the anisotropies detected and their statistical significance compared to other models; in Section~\ref{sec:mocks} we investigate our ability to distinguish anisotropic $H_0$ and bulk flows for current and future Tully-Fisher datasets; finally, in Section~\ref{sec:conclusions} we present the conclusions of this work.

\section{Data}
\label{sec:data}

\begin{figure*}[t]
\centering
\includegraphics[width=0.8\textwidth]{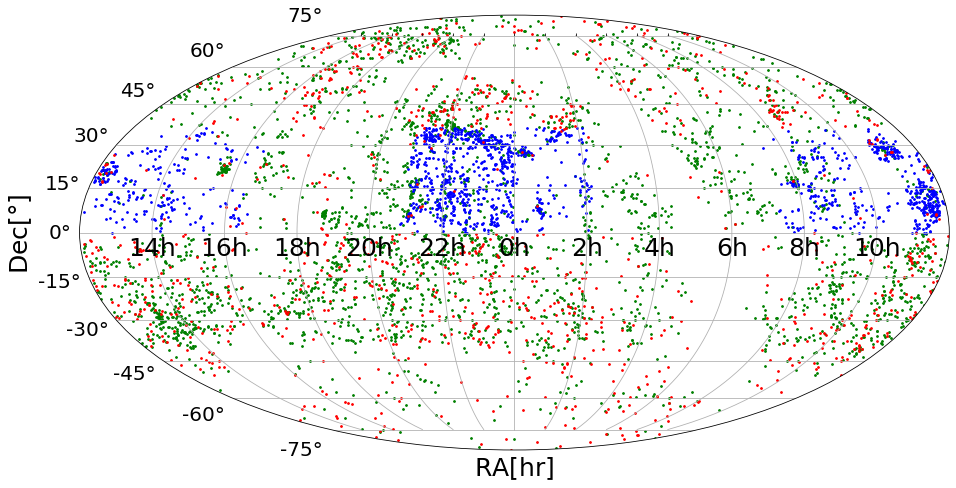}
\caption{Sky distribution of \HI\ sources with WISE magnitudes from the CF4 Tully-Fisher catalog. In blue, 1424 \HI\ line-width measurements from ALFALFA \cite{2018alf}; in green, 2979 non-ALFALFA \HI\ line-width measurements from the ADHI catalog \cite{Courtois_2009}; in red, 1076 other sources from the \textit{Springob/Cornell \HI\ catalog} \cite{Springob_2005} or the \textit{Pre Digital \HI\ catalog} in EDD.}
\label{fig:logW_sky}
\end{figure*}

The \textit{Cosmicflows-4} (CF4) catalog \cite{2020cf4} is currently the largest full-sky catalog of galaxies with Tully-Fisher distances and peculiar velocities. It is derived from heterogeneous datasets and contains 10,737 galaxies with \HI\ redshifts and line widths, together with optical or infrared photometry. Because the method of this paper relies on the identification of physical differences in the Tully-Fisher relationship in different regions of the sky, it is highly sensitive to systematic differences between sources of photometry or \HI\ line widths. We therefore want the data to be as uniform as possible.

\subsection{Photometry}

Although $i$-band optical photometry from the \textit{Sloan Digital Sky Survey} \cite[SDSS;][]{York_2000} is available for 7502 CF4 galaxies in the northern sky, we choose to use $W1$-band infrared photometry from the all-sky \textit{Wide-field Infrared Satellite Explorer} \cite[WISE;][]{Wright_2010}, available for 5479 CF4 galaxies, as a single source of photometry over the whole sky mitigates systematic variations.

\subsection{\texorpdfstring{\HI}{HI} line widths}

In contrast to photometry, no homogeneous all-sky \HI\ dataset exists. \textit{Cosmicflows-4} \HI\ line widths are therefore compiled from a variety of sources, which are converted to a standard quantity, $\Wmxc$ \cite{Kourkchi_2019}. The \HI\ data used here are taken primarily from the \textit{All Digital \HI\ (ADHI) catalog} \cite{Courtois_2009}, which is mainly composed of good quality \HI\ line widths from the ALFALFA survey \cite{2018alf}. Sources not covered by the ALFALFA survey have \HI\ line widths from the \textit{Springob/Cornell \HI\ catalog} \cite{Springob_2005} or the \textit{Pre Digital \HI\ catalog} on EDD\footnote{http://edd.ifa.hawaii.edu; `Pre Digital HI'}, both containing measurements from a variety of large single-dish radio telescopes. The sky distributions of these various sources of \HI\ line widths are shown in Figure~\ref{fig:logW_sky}. 

\begin{figure}[t]
\centering
\includegraphics[width=0.5\textwidth]{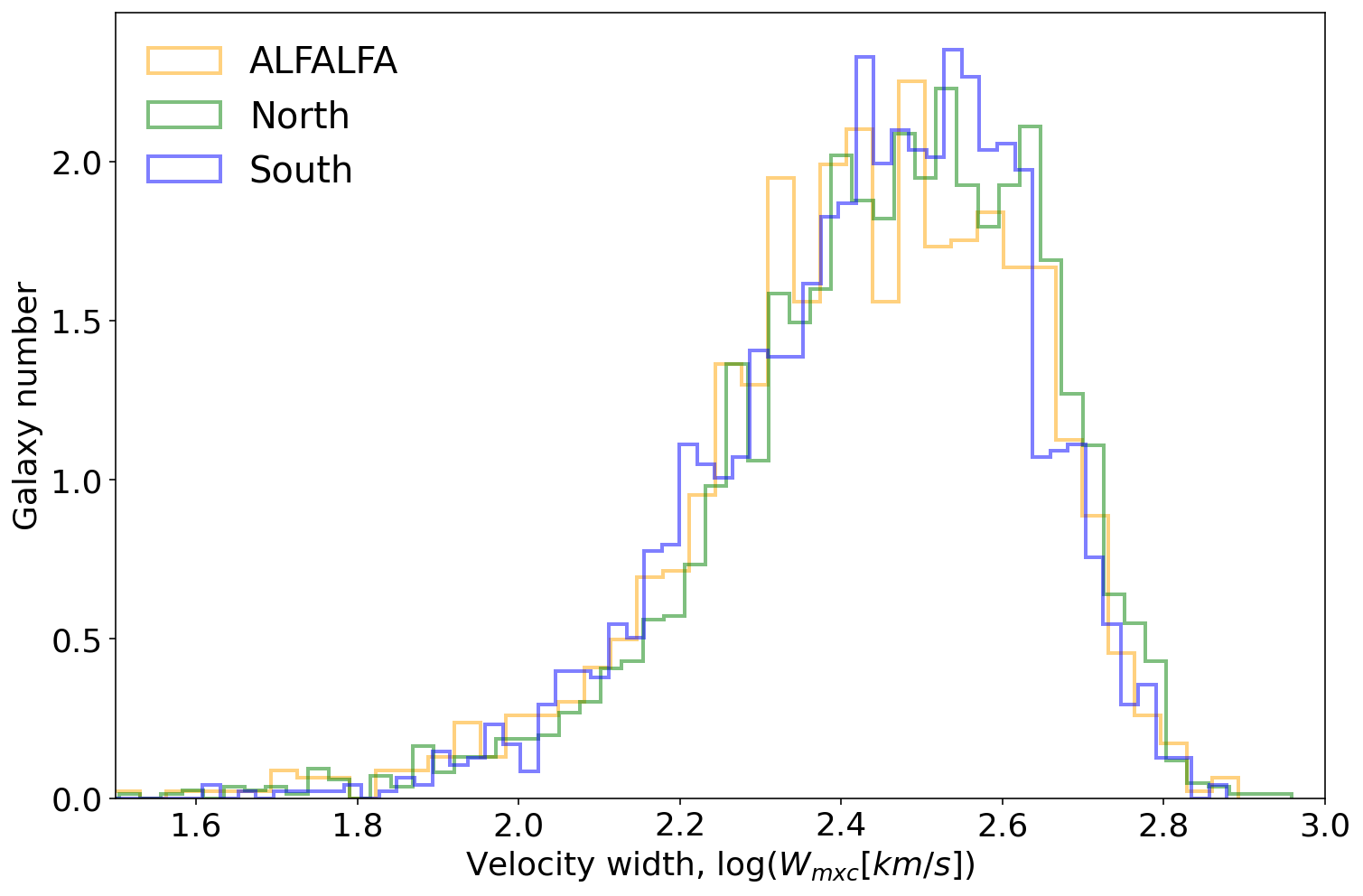}
\caption{Normalized distributions of \HI\ velocity widths from the CF4 Tully-Fisher catalog for all sources with $W1$ magnitude measurements from WISE. The distribution for sources in the northern sky are shown in green, sources in the southern sky in blue, and sources with line-width measurements from ALFALFA in orange.}
\label{fig:logW}
\end{figure}

Even though the WISE data are all-sky, the 1424 sources covered by ALFALFA are all in the northern sky. If our analysis only used this single source of \HI\ data, the sky coverage would be too limited to meaningfully probe anisotropy. We therefore use the full \textit{Cosmicflows-4} \HI\ dataset and check that there are no remaining systematic differences in \HI\ measurements between the different sources and hemispheres. In Figure~\ref{fig:logW} we compare the distributions of line widths from ALFALFA, from measurements in the northern sky, and measurements in the southern sky. The median values and standard deviations of the distributions are $\log{\Wmxc}=2.45 \pm 0.18$ in the north and $\log{\Wmxc}=2.46 \pm 0.17$ in the south. Comparing the northern sky with the southern sky, a two-sample Kolmogorov–Smirnov (KS) test gives us a test statistic of 0.034 with a $p$-value of 0.077, providing no evidence the distributions differ significantly. Since it is not likely that an all-sky \HI\ dataset will exist in the near future, there will continue to be a need to ensure that north-south \HI\ measurement methodologies are consistent. This will remain a limitation of this analysis even with future datasets. A systematic difference as small as $\Delta \log{\Wmxc} = 0.01$ corresponds roughly to a 3\% effect on $H_0$.

\section{Method}
\label{sec:method}

We have developed a forward-modeling methodology for simultaneously fitting the Tully-Fisher relation and peculiar velocity field in a sample of galaxies and applied it to the CF4 Tully-Fisher data \cite{Boubel_2024} (see also \cite{Kourkchi_2020}). The method formulates the conditional probability of observing an apparent magnitude $m$ for a galaxy as
\begin{equation}
P(m\,|\,w,z,\alpha,\delta,\parTF,\parPV) = \frac{F(m)\exp\left[-\frac{(m-m^\prime)^{2}}{2\sigma_\textrm{TF}^2}\right]}{\int F(m)\exp\left[-\frac{(m-m^\prime)^{2}}{2\sigma_\textrm{TF}^2}\right]\,dm}
\label{eqn:mcondprob}
\end{equation}
where $m^\prime$ is the predicted apparent magnitude as a function of the observed quantities \HI\ velocity width $w$, redshift $z$, and position ($\alpha$,$\delta$), and the parameters of the models for the Tully-Fisher relation $\theta_{\rm TF}$ and the peculiar velocity field $\theta_{\rm PV}$,
\begin{equation}
m^\prime(w,z,\alpha,\delta,\parPV,\parTF) = 
M^\prime(w) + 25 + 5\log(1+z) + 5\log d_C(z_c^\prime) ~.
\label{eqn:mpred}
\end{equation}
This predicted apparent magnitude is given in terms of $M^\prime(w)$, the predicted absolute magnitude from the Tully-Fisher relation model for the observed \HI\ velocity width $w \equiv \log{\Wmxc} - 2.5$, and $z^\prime_c$, the predicted co-moving redshift from the peculiar velocity model for the observed redshift and position.

The zeropoint of the Tully-Fisher relation and the value of $h = H_0/100$ are directly related \cite{Boubel_2024b}; specifically, a shift in $5 \log h$ corresponds to a shift in $M$ given by
\begin{equation} 
M(w,h) = M(w,h=1) + 5 \log h ~.
\label{eqn:abs_mag_h}
\end{equation}
 
Consequently, it is not possible to use the Tully-Fisher relation on its own to determine $H_0$. However, it is in principle possible to detect variations in $H_0$ using the \emph{differential} Tully-Fisher relation, since changes in $H_0$  would be reflected by shifts in the Tully-Fisher zeropoint. This application does not rely on primary distance calibrations, but it does require high confidence in the spatial uniformity of the Tully-Fisher measurements, with negligible position-related systematic errors in the data. 

Eq.~\ref{eqn:abs_mag_h} implies that a positive shift in $M(w)$, i.e. shifting the Tully-Fisher relation towards fainter magnitudes, corresponds to a positive shift in $h$. Intuitively, fainter predicted absolute magnitudes necessitate closer inferred distances if the apparent magnitude is held fixed, and so a larger $H_0$ is inferred for a fixed observed redshift. As a result, a positive Tully-Fisher zeropoint anisotropy corresponds to a positive $H_0$ anisotropy. 

This paper explores the extent to which it is possible to measure $H_0$ anisotropies using the Tully-Fisher relation using current and future datasets.

We can model a direction-dependent $H_0$ by allowing the differential Tully-Fisher zeropoint to vary across the sky. In this scenario, the Tully-Fisher model \cite{Boubel_2024} for the absolute magnitude given the velocity width, $M^\prime(w)$, would have the form
\begin{equation} 
M =
\begin{cases}
a_0(\ell,b) + a_1w & (w < 0) \\
a_0(\ell,b) + a_1w + a_2w^2 & (w \geq 0) \\
\end{cases}
\label{eqn:TFreln}
\end{equation}
where we have allowed $a_0$ to be a function of position on the sky, specified by $\ell$ (Galactic longitude) and $b$ (Galactic latitude). Representing the variation on the sky in terms of a series expansion, we can fit the lowest-order spherical multipoles. The monopole and dipole terms can described using four parameters ($a_{00}$, $a_{0x}$, $a_{0y}$, $a_{0z}$) as
\begin{equation}
a_0(\ell,b) = a_{00} + \widetilde{a_0}(\ell,b) = a_{00} + a_{0x}\cos{(b)}\cos{(\ell)} - a_{0y}\cos{(b)}\sin{(\ell)} + a_{0z}\sin{(b)}
\label{eqn:zp_dipole}
\end{equation}
where the components of the dipole ($a_{0x}$, $a_{0y}$, $a_{0z}$) are represented in Cartesian coordinates in the Galactic reference frame in order to have Gaussian-distributed parameters. 

In this work, the anisotropy model will be denoted $\widetilde{a_0}(\ell,b)$ and will either be truncated at the dipole term, as above, or the quadrupole term, adding 5 more free parameters. Because the redshift range of the CF4 data is very limited, there is no need to include a decay factor as a function of redshift \cite{Cowell_2023, Macpherson_2021}, although this may be a consideration for future studies with more extensive datasets.

Assuming that the intrinsic Tully-Fisher relation is the same everywhere and that there are no differences in photometric calibration between different regions of the sky, variations in $a_0$ on the sky are due to variations in $H_0$ (parametrized as $h = H_0/100$\,\kmsMpc). The deviation of the measured Tully-Fisher zeropoint from its true value is
\begin{equation}
M(w,h(\alpha,\delta)) - M(w,\bar{h}) = \widetilde{a_0} = 5\log h(\alpha,\delta) - 5 \log \bar{h} 
= 5 \log (\Delta h(\alpha,\delta) / \bar{h} + 1),
\end{equation}
where $\bar{h}$ is the mean value of $h$. This leads to 
\begin{equation}
\Delta H_0 = \overline{H_0} (10^{\widetilde{a_0}/5} - 1),
\label{eq:deltaH}
\end{equation}
where we have replaced $h$ with $H_0$. Thus, differences in $H_0$ on the sky can be directly linked to the anisotropy of the Tully-Fisher zeropoint, $\widetilde{a_0}$. 

In general, the Tully-Fisher parameters $a_0$, $a_1$, $a_2$, $\epsilon_0$, and $\epsilon_1$ (see ~\cite{Boubel_2024}) are unique to the dataset, as the Tully-Fisher relation changes depending on the photometric band. However, any real $H_0$ anisotropy should result in a consistent $\widetilde{a_0}$ across datasets. We can therefore, in principle, combine datasets to achieve greater precision on constraints for the multipole terms of $\widetilde{a_0}$, whilst allowing the other Tully-Fisher parameters to vary. This only works if there are no spatial non-uniformities within individual photometric datasets.

\section{Results}
\label{sec:results}

In this work, we use the $W1$-band Tully-Fisher data from CF4 because it provides uniform all-sky photometry. We apply a lower redshift limit, requiring $cz > 3000$\,\kms, as the large relative effects of peculiar velocities at low redshifts may have a significant impact on $H_0$ determinations. This is the same limit as was chosen in ~\cite{Boubel_2024b} to produce a $H_0$ that did not vary with redshift. Higher redshift limits reduced the size of the sample while only resulting in small changes ($< 1\sigma$) to the amplitude of the dipole; lower redshift limits slightly increased the significance of the dipole amplitude. Because a lower redshift limit greater than or equal to 3000\,\kms\ produced consistent results, this value was chosen to preserve sample size while minimizing the effects of peculiar velocities.

\subsection{\texorpdfstring{$H_0$}{H0} dipole and quadrupole}
\label{sec:dipole+quadrupole}

\begin{figure*}
\centering
\begin{subfigure}{0.8\textwidth}
  \centering
  \includegraphics[width=1\textwidth]{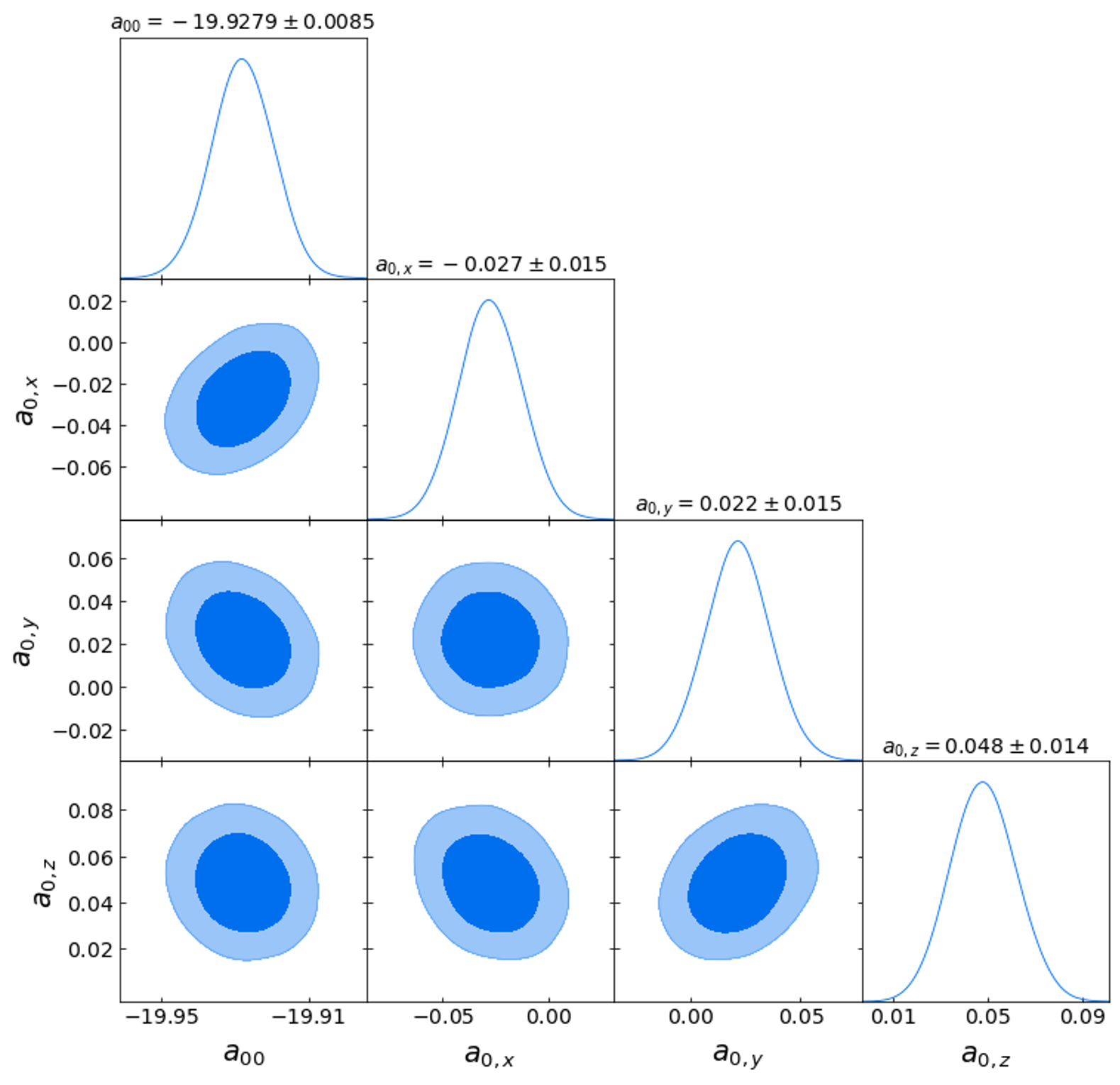}
  \label{fig:cf4_cart}
\end{subfigure}
\begin{subfigure}{0.8\textwidth}
  \centering
  \includegraphics[width=1\linewidth]{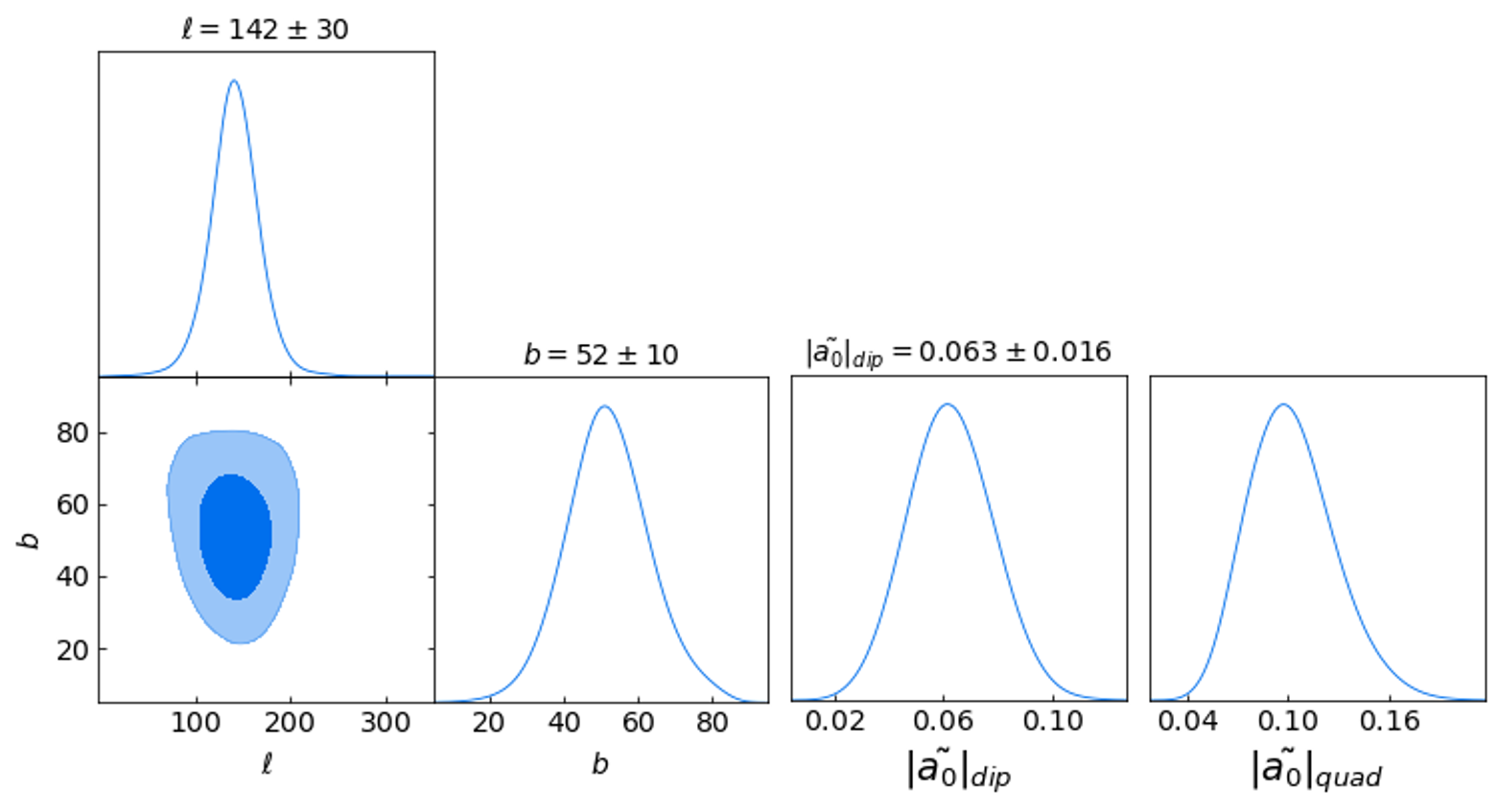}
  \label{fig:cf4_sph}
\end{subfigure}
\caption{Constraints from fitting a Tully-Fisher zeropoint monopole, dipole, and quadrupole to the $W1$ CF4 data. Top cornerplot: best-fit monopole $a_{00}$=$-$19.928$\pm$0.009\,mag and dipole $\widetilde{a_0}$ = ($a_{0x}$, $a_{0y}$, $a_{0z}$) = ($-$0.027\,$\pm$\,0.015, 0.022\,$\pm$\,0.015, 0.048\,$\pm$\,0.014)\,mag. Bottom cornerplot: dipole direction is ($\ell$,$b$) = (142\,$\pm$\,30\degree,52\,$\pm$\,10\degree) and amplitude is $|\widetilde{a_0}|_\textrm{dip}$=0.063\,$\pm$\,0.016\,mag; best-fit quadrupole amplitude is $|\widetilde{a_0}|_{\textrm{quad}}$\,=\,0.09\,$\pm$\,0.08\,mag.}
\label{fig:cf4_constraints}
\end{figure*}

\begin{figure*}
\centering
\includegraphics[width=0.55\textwidth]{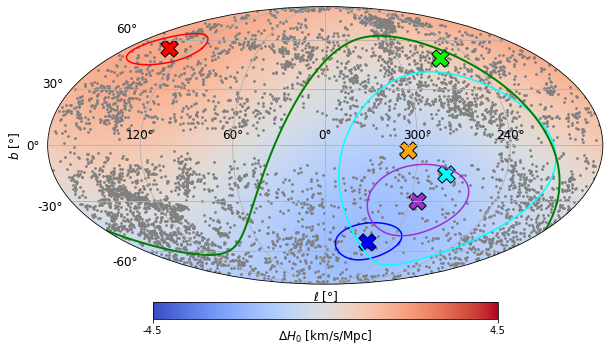}
\includegraphics[width=0.55\textwidth]{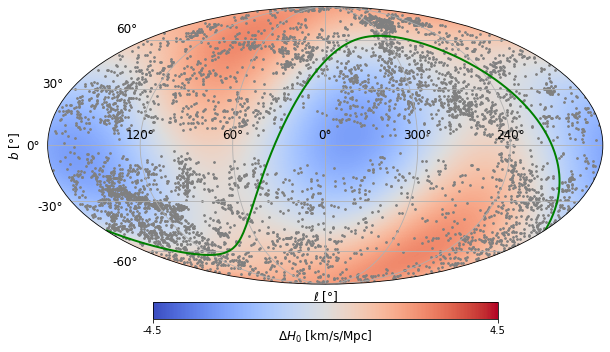}
\includegraphics[width=0.55\textwidth]{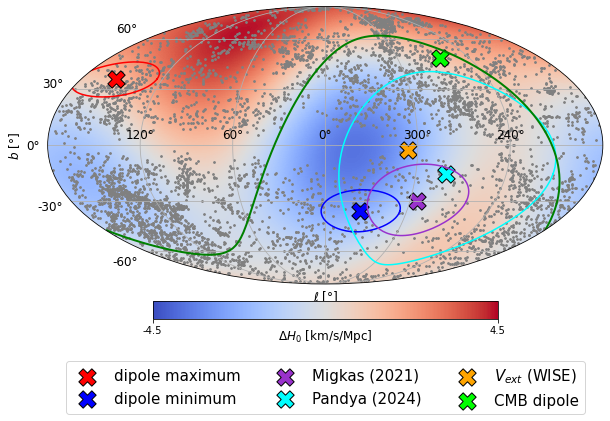}
\caption{Mollweide projections of anisotropies in Galactic coordinates. Top: The best-fit $H_0$ dipole, where the direction of its +/$-$ value is shown by the red/blue crosses and the 1$\sigma$ error boundary is shown by the red/blue ellipses. The amplitude is $\Delta H_0$ = 2.10\,$\pm$\,0.53\,\kmsMpc\ if $H_0$ = 70\,\kmsMpc and the direction of the maximum is ($\ell, b$) = (142\,$\pm$\,30\degree, 52\,$\pm$\,10\degree). The gray points are the $W1$-band CF4 galaxies used in this fitting. The direction of the \textit{Planck} CMB dipole \cite{2020_Planck} in the heliocentric frame is shown by the green cross. The direction of the external bulk flow from outside the 2M++ volume $\mathbf{V}_{\textrm{ext}}$ ($W1$-band fit from ~\cite{Boubel_2024}) is shown by the orange cross. The purple and light blue crosses and ellipses are directions of maximum $H_0$ anisotropy measured by ~\cite{Migkas_2021, Pandya_2024} using galaxy scaling relations. The green line traces out the celestial equator. Middle: Best-fit $H_0$ quadrupole with amplitude $\Delta H_0$ = 3.0\,$\pm$\,2.6\,\kmsMpc\ if $H_0$ = 70\kmsMpc. Bottom: Sum of best-fit dipole and quadrupole.}
\label{fig:w1_dipole}
\end{figure*}

Figure~\ref{fig:cf4_constraints} shows the pairwise constraints, with contours at the 68\% and 95\% confidence levels, from fitting an $a_0$ monopole, dipole, and quadrupole to the CF4 $W1$-band data. The dipole is measured to have a significance of 3.9$\sigma$. Figure~\ref{fig:w1_dipole} is a visualization of these $a_0$ anisotropies on the sky, converting $a_0$ to the corresponding $\Delta H_0$ assuming $H_0$\,=\,70\,\kmsMpc. The top panel of Figure~\ref{fig:cf4_constraints} shows the fitted dipole, amplitude $\Delta H_0$\,=\,2.10\,$\pm$\,0.53\,\kmsMpc\ in the direction ($\ell, b$) = (142\,$\pm$\,30\degree, 52\,$\pm$\,10\degree). The dipole direction is not aligned with the external bulk flow fitted from the same data \cite{Boubel_2024} nor with the CMB dipole determined by \textit{Planck} \cite{2020_Planck}. The dipole minimum, however, is consistent with the direction of the largest (negative) $H_0$ anisotropy found in studies using galaxy cluster scaling relations ~\cite{Migkas_2021, Pandya_2024}. This minimum occurs at the antipode, ($\ell, b$) = (322\,$\pm$\,29\degree, $-$52\,$\pm$\,12\degree).

We also fit a combination of a dipole and a quadrupole; the quadrupole adds 5 free parameters to the model. The direction is not fixed, as it was in previous studies \cite{Sah_2024, Cowell_2023}. The middle panel of Figure~\ref{fig:w1_dipole} shows the best-fitting quadrupole (only); it has an amplitude of $|\widetilde{a_0}|$\,=\,0.09\,$\pm$\,0.08\,mag ($\Delta H_0$\,=\,3.0\,$\pm$\,2.6\,\kmsMpc\ if $H_0$\,=\,70\,\kmsMpc). The significance of this quadrupole term is low, only 1.1$\sigma$. The bottom panel of the figure shows the combined best-fit dipole plus quadrupole model. 

\subsection{Comparison with a bulk flow model}
\label{sec:compare}

The presence of a residual bulk flow in the sample could be misinterpreted as an anisotropy in $a_0$, because galaxy peculiar velocities modify their observed redshifts and so affect the predicted absolute magnitudes. As with $H_0$, the direction of the bulk flow is in the same direction as that of the $a_0$ dipole maximum. This can be understood by looking at the effect of a peculiar velocity $v$ on the inferred cosmological redshift $z_c$ at a fixed observed redshift $z$:
\begin{equation}
1 + z = (1+z_c)(1+v/c).
\end{equation}
A higher, positive value for $v$ reduces $z_c$, resulting in a smaller inferred distance (for fixed $H_0$). A closer galaxy must therefore be intrinsically fainter, since its apparent magnitude is a fixed observed quantity. Thus the Tully-Fisher zeropoint is shifted to more positive values.

However, a bulk flow is, in principle, distinguishable from a $H_0$ sky variation, mainly because the effect on $a_0$ of a bulk flow depends on a galaxy's redshift while the effect of a $H_0$ variation does not. As a result, a bulk flow will not create a pure spatial dipole in $a_0$ unless the redshift is fixed. Thus, redshift coverage and sky coverage are both important to being able to distinguish between them. 

For the currently available data, the CF4 Tully-Fisher catalog, we can fit a zeropoint dipole and quadrupole, as described in the previous sections, or we can instead fit a velocity dipole, or we can try to constrain both simultaneously. To compare the fits from these models, we can compute the Bayes factor
\begin{equation}
B_{01} \equiv \frac{P(d|M_0)}{P(d|M_1)}
\end{equation}
where $B_{01}$ is the posterior odds that model $M_0$ is true rather than model $M_1$, in light of the data $d$ and assuming equal priors for both models \cite{trotta_2008, liddle_2009}. Here, the number of data points $n$ is much greater than the number of parameters $k$, so we can use the approximation
\begin{equation}
P(d|M) \approx \exp(-(k\ln{n}-2\ln{\hat{L}})/2)
\end{equation}
where $\hat{L}$ is the maximum likelihood of the model $M$. The quantity $k\ln{n} - 2\ln{\hat{L}}$ is the Bayesian information criterion, $\textrm{BIC}$ \cite{schwarz_1978}. A lower $\textrm{BIC}$ is preferred, meaning that a model is penalized if it requires more parameters $k$. So,
\begin{equation}
\ln (B_{01}) \approx  (\textrm{BIC}(M_1) - \textrm{BIC}(M_0))/2 ~.
\end{equation}

Table~\ref{tab:lnB} lists the Bayes factors for different model fits to the CF4 $W1$ Tully-Fisher data. Using the recommended interpretation of ~\cite{kass_1995} (adapted from the original Jeffreys scale \cite{jeffreys_1998}), also used in other cosmological analyses (for example: refs.~\cite{john_2002, drell_2000}), we see strong evidence of a velocity dipole rather than a $H_0$ dipole if we had to choose between the two. Secondly, there is weak evidence that a velocity dipole with an additional $H_0$ dipole is favored over a velocity dipole only. Lastly, a $H_0$ dipole term alone is strongly favored over a model that incorporates both a dipole and a quadrupole, suggesting that the quadrupole term adds too many parameters with minimal improvement in fitting the data.

\begin{table}[t]
\centering
\begin{tabular}{|l|l|c|}
\hline
$M_0$ & $M_1$ & $\ln(B_{01})$ \\
\hline
velocity dipole & $H_0$ dipole & 4.7 \\
$H_0$ dipole & $H_0$ dipole and quadrupole & 7.5 \\
velocity and $H_0$ dipole & $H_0$ dipole & 5.7 \\
velocity dipole and $H_0$ dipole & velocity & 0.99 \\
\hline
\end{tabular}
\caption{Bayes factors of fitting various models to the CF4 $W1$ Tully-Fisher data. In each scenario, $M_0$ is the preferred model with the lower BIC. The third column is the Bayes factor, $\ln({B_{01}})$, and a higher value indicates stronger evidence for the preferred model.\label{tab:lnB}}
\end{table}

\begin{figure}[t]
\centering
\begin{subfigure}{0.7\textwidth}
  \centering
    \caption{$H_0$ anisotropy.}
  \includegraphics[width=1\textwidth]{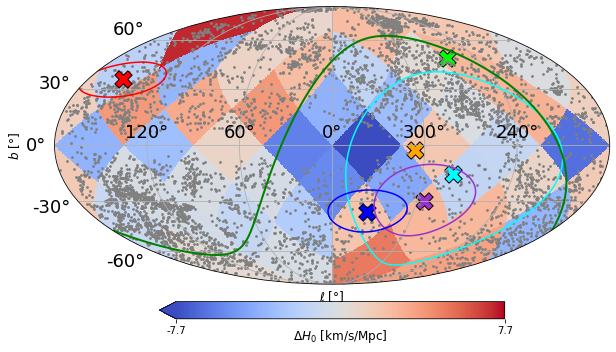}
  \label{fig:hp_deltaH}
\end{subfigure}
\begin{subfigure}{0.7\textwidth}
  \centering
    \caption{Standard errors.}
  \includegraphics[width=1\textwidth]{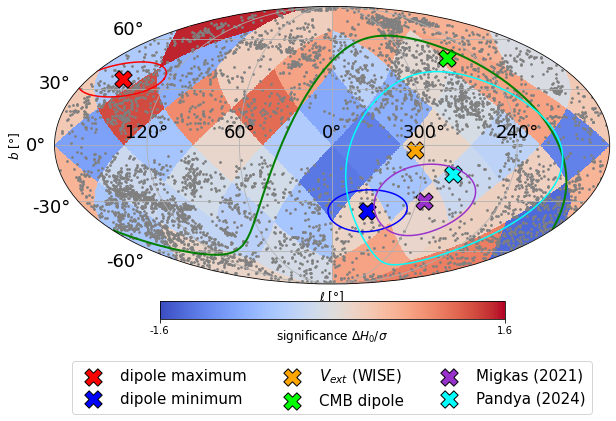}
  \label{fig:hp_sigma}
\end{subfigure}
\caption{(a)~Best-fit Tully-Fisher zeropoints for CF4 $W1$ data (gray points) binned in HEALPix pixels (resolution $\textrm{NSIDE}$\,=\,2), expressed as $\Delta H_0$ using Equation~\eqref{eq:deltaH}. The red/blue crosses show the best-fit +/$-$ dipole direction determined in Section~\ref{sec:dipole+quadrupole}. (b)~Significance of the anisotropy in each pixel, as indicated by the best-fit values from the top panel divided by their standard errors.}
\label{fig:binned_a0}
\end{figure}

\subsection{Sample variance}

The scale of homogeneity is usually taken to be $\sim$100\hMpc\ \citep{Scrimgeour_2012, Martinez_1998}, roughly corresponding to a sample volume with a redshift limit  $z>0.03$. The current \textit{Cosmicflows-4} Tully-Fisher sample has an effective redshift of only $z_{\textrm{eff}}=0.017$, where local density fluctuations are significant, and so it cannot be taken to be a statistically representative volume of the Universe. In order to fairly test the homogeneity and isotropy, the larger volumes provided by surveys like WALLABY and DESI (with effective redshifts $z_{\textrm{eff}}>0.05$) will be essential. 

To highlight this point, we can compare the sample variance of the local $H_0$ value expected from $\Lambda$CDM in simulations using different effective depths. A few studies have performed these calculations \citep{Camarena_2018, Zhai_2024, Wojtak_2013} and their results are similar: for the redshift range of the \textit{Cosmicflows-4} dataset, the variance in $H_0$ is typically $\sim$1.3\kmsMpc, whereas for a survey out to $z=0.1$, it is $\sim$0.6\kmsMpc. Although this does not explain the Hubble tension \citep{Camarena_2022, Wu_2017}, it does decrease its significance, and will do the same for any $H_0$ anisotropy detection. For future applications of the method described here, a proper characterisation of sample variance in the local $H_0$ value and its higher-order anisotropies will be essential.

There is also a second way sample variance plays a role in this analysis. The local bulk motion, which should be consistent with the sample variance in the velocity field, depends on the scale of the survey. Ref.~\cite{Whitford_2023} determined that, at an effective depth of $\sim$70\,Mpc (i.e.\ for the \textit{Cosmicflows-4} Tully-Fisher sample), the $\Lambda$CDM prediction for the rms bulk flow is roughly 300\,$\pm$\,100\kms, depending on survey geometry. The measured bulk flows for this volume are generally consistent with this prediction \citep{Carrick_2015, Boubel_2024}. Such a bulk flow is difficult to distinguish from an $H_0$ dipole (see Section~\ref{sec:compare}), but in a larger survey the bulk flow should be smaller while the effect of an $H_0$ dipole would be greater. At an effective depth of $\sim$200\,Mpc (i.e.\ for the WALLABY/DESI Tully-Fisher samples), the bulk flow is expected to be only $\sim$100\,$\pm$\,50\kms\ \cite{Whitford_2023}. However, since many studies find the measured bulk flow at these greater effective depths to be in tension with $\Lambda$CDM \citep{Whitford_2023}, we will conservatively assume the bulk flow from the smaller volume even in our forecasts for future surveys (Section~\ref{sec:futuremocks}).  

\subsection{Pixel map of anisotropies}

In the previous sections, we have fit for the dipole and quadrupole variation of $H_0$ on the sky. We lack sufficient data to extract higher-order information with confidence (even a quadrupole adds little or no value; see Table~\ref{tab:lnB}). However, we can get a rough picture of any smaller-scale anisotropies, if they exist, by binning the data on the sky and fitting a constant $a_0$ separately for each bin.

For this exercise, we choose the HEALPix \cite{Gorski_2005} binning scheme with $\textrm{NSIDE}$\,=\,2, corresponding to 48 sky pixels of equal area. Using the CF4 $W1$ data, we fit $a_0$ for the galaxies in each pixel, fixing all other Tully-Fisher and peculiar velocity parameters to the fitted values from the full dataset. To find the mean, $a_{00}$, we take the average of the $a_0$ values weighted by uncertainty. Then, $a_{0} - a_{00}$ gives the value of $\widetilde{a_0}(\alpha,\delta)$ for that pixel, which can be converted to $\Delta H_0$ using Equation~\eqref{eq:deltaH}. The result is the map shown in Figure~\ref{fig:binned_a0}.

\section{Simulations}
\label{sec:mocks}

To estimate the statistical uncertainties on the dipole measurement, $\widetilde{a_0}$, we simulate mock CF4 datasets. The procedure for generating these mocks is detailed in Section~5 of \cite{Boubel_2024}, but we have modified this slightly as described below. We note that these mocks reproduce the observed local structures and statistics of CF4. 

\subsection{Bulk flow versus \texorpdfstring{$H_0$}{H0} dipole}
\label{sec:cf4mocks}

In our previous work \cite{Boubel_2024}, the external bulk flow for the CF4 Tully-Fisher dataset was separately determined for the $i$ and $W1$ bands. In Cartesian equatorial coordinates, the external velocity dipole in the $W1$-band was found to be $\mathbf{V}_{\textrm{ext}}$\,=\,($-$90\,$\pm$\,10,+35\,$\pm$\,9,$-$209\,$\pm$\,9)\kms. If we assume such a velocity dipole is present in the data, we can show if and how it can be distinguished from a true $a_0$ (and corresponding $H_0$) dipole using mocks.

We use the procedure of ~\cite{Boubel_2024} to create our mocks, but without peculiar velocities. We add a bulk flow of $\mathbf{V}_{\textrm{ext}}$\,=\,($-$90\,$\pm$\,10,+35\,$\pm$\,9,$-$209\,$\pm$\,9)\kms\ to each galaxy, modifying its observed redshift and thus its apparent magnitude (see Equation~\eqref{eqn:mpred}). Fitting the Tully-Fisher relation and a bulk flow term to these mocks should faithfully recover all parameters. Alternatively, fitting a Tully-Fisher relation with a zeropoint dipole and no bulk flow term gives a model for an $H_0$ dipole that approximates the actual bulk flow.

\begin{figure}[H]
\centering
\begin{subfigure}{0.85\textwidth}
  \centering  \includegraphics[width=1\textwidth]{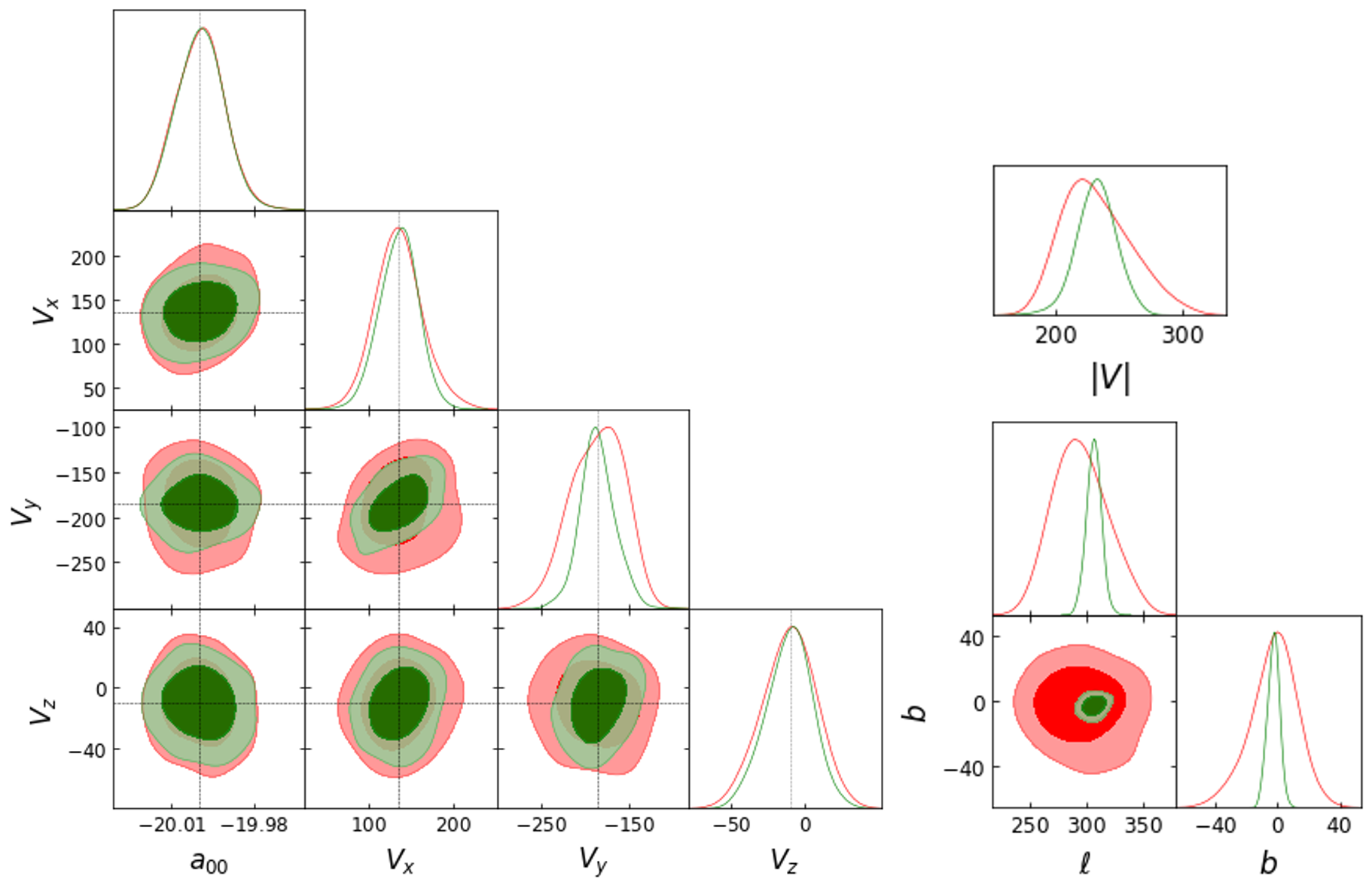}
      \caption{Fitting a velocity bulk flow.}
  \label{fig:vp_fit_mocks}
\end{subfigure} \\
\begin{subfigure}{0.85\textwidth}
  \centering  \includegraphics[width=1\textwidth]{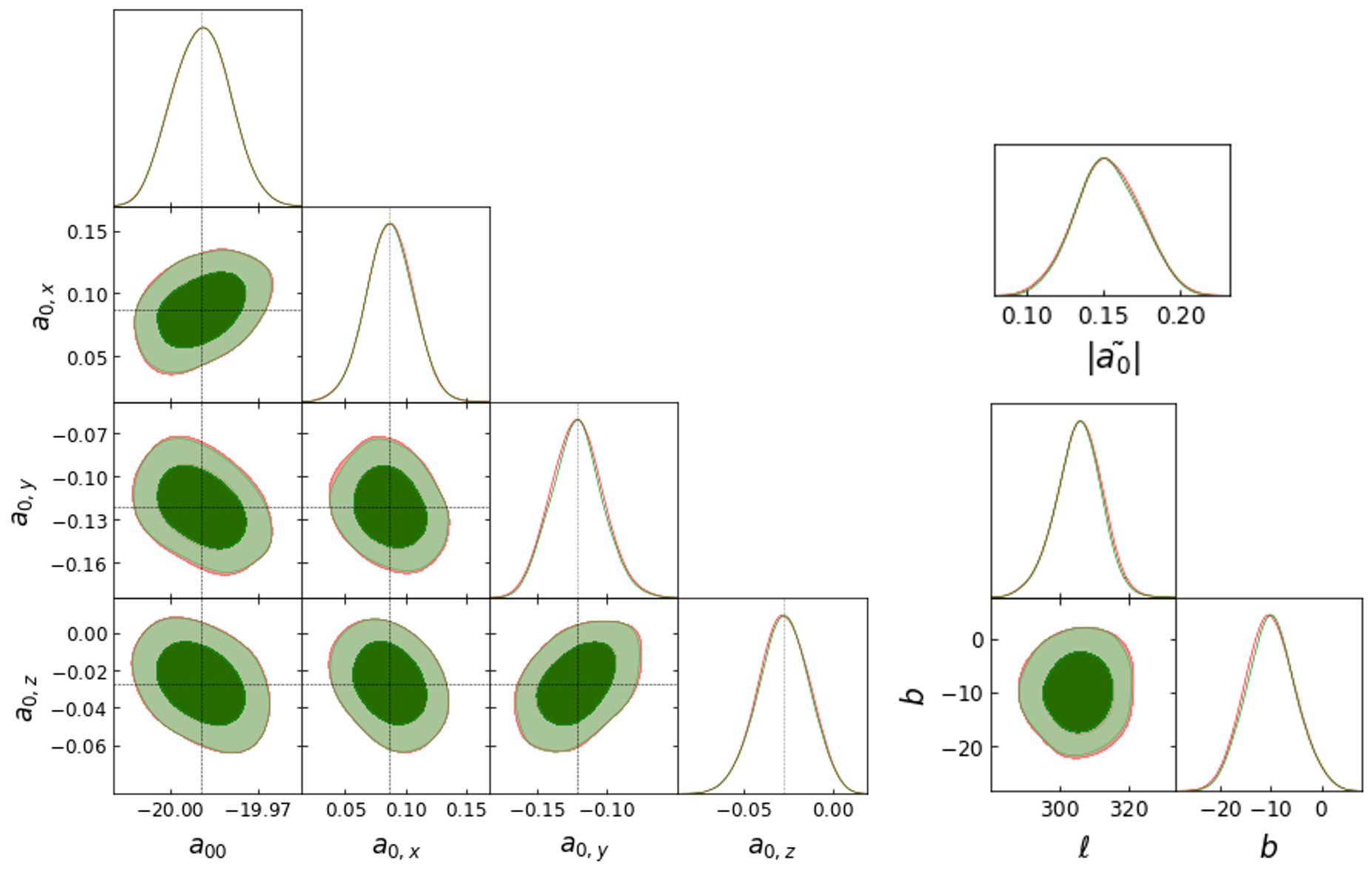}
      \caption{Fitting a Tully-Fisher zeropoint dipole.}
  \label{fig:zp_fit_mocks}
\end{subfigure} \\
\caption{(a)~Fitting a velocity dipole ($V_x$, $V_y$, $V_z$) in \kms\ to 1000 CF4 mocks. The mocks shown in green used a velocity dipole indicated by the gray lines; the mocks shown in red used an approximately corresponding $H_0$ dipole. (b)~Fitting a $H_0$ dipole model (Eq.\eqref{eqn:zp_dipole}) to 1000 CF4 mocks. The mocks shown in green used a $H_0$ dipole indicated by the gray lines; the mocks in red used an approximately corresponding velocity field dipole. In both panels, the 2D distributions are shown on the left for the Cartesian components and the same information is presented in spherical coordinates on the right: 1D distributions for the dipole amplitude of (a)~$\mathbf{V}_{\textrm{ext}}$ in \kms\ or (b)~$\widetilde{a_0}$ in mag, plus 2D distributions for the dipole direction in Galactic coordinates $\ell$ and $b$ in degrees.}
\label{fig:mock_dipole}
\end{figure}

This model can then be used to generate a new suite of mocks with a zeropoint ($H_0$) dipole rather than a bulk flow. Again, we fit both a velocity dipole and a zeropoint dipole to this new set of mocks. This exercise provides insight into our ability to distinguish a $H_0$ dipole from a velocity dipole in the existing dataset. Figure~\ref{fig:mock_dipole} shows the resulting pairwise constraints on $a_{00}$, $a_{0x}$, $a_{0y}$, and $a_{0z}$ for each of these four tests. Green contours represent cases where the fitted model matches the simulated model, while red contours represent fits of the incorrect model. The gray hashed lines show the input parameters of the simulations. 

The power of the CF4 Tully-Fisher dataset to determine which is the correct dipole model is summarized in the left panel of Figure~\ref{fig:bayes_future}. This shows the distribution of Bayes factors favoring the correct model over the incorrect model in mock CF4 datasets. If there is an $H_0$ (i.e., $a_0$) dipole in the CF4 Tully-Fisher data, we would not be able to easily determine if we have the correct model or not, as the Bayes factor is relatively small (blue histogram; median $\ln(B_{01}) = 2$). However, if there is a velocity dipole (bulk flow) in the CF4 Tully-Fisher data, the Bayes factor would show significant evidence favoring the correct model (red histogram; median $\ln(B_{01}) = 79$).

There are two reasons for this. First, the intrinsic scatter in the Tully-Fisher relation for CF4 data is at least 0.5~mag \cite{Boubel_2024}, corresponding to about a 25\% error on $H_0$ that depends on distance, whereas the bulk flow has a an error of only $\pm$\,10\kms\ in each component \cite{Boubel_2024}, corresponding to a $\sim$0.1\% variation in $H_0$ that is distance-independent. Second, the direction we have chosen for the dipoles lies close to the Zone of Avoidance (the Milky Way plane), so the dipole is less constrained given the sky distribution of our mock sample. 

\subsection{Forecasts for future datasets}

CF4 is currently the largest collection of galaxies with Tully-Fisher data, but this will change rapidly with upcoming \HI\ surveys like the \textit{Wide-field ASKAP L-band Legacy All-sky Blind surveY} \cite[WALLABY;][]{Koribalski_2020, Westmeier_2022} and the \textit{FAST All Sky \HI\ Survey} \cite[FASHI;][]{Kang_2022, Zhang_2023}. FASHI is expected to detect more than 100,000 \HI\ sources covering 22,000\,deg$^2$ between declinations $-$14\degree and +66\degree \cite{Kang_2022}. WALLABY is expected to detect 210,000 \HI\ sources over 14,000\,deg$^2$ \cite{Westmeier_2022} of the southern sky. Both surveys have a redshift limit around $z\approx0.1$. Not all these \HI\ sources can be used in fitting the Tully-Fisher relation, as the signal-to-noise ratio needs to be sufficiently high and the inclination sufficiently edge-on. In WALLABY, about 40\% of sources are expected to meet these criteria \cite{Courtois_2022}. Additionally, the DESI peculiar velocity survey \cite[DESI;][]{Saudler_2023} is expected to obtain stellar rotation velocities enabling about 50,000 Tully-Fisher peculiar velocity measurements over 14,000\,deg$^2$ of the northern sky. These surveys should result in an order-of-magnitude increase in sample size relative to the CF4 Tully-Fisher catalog over the next few years. They will also reach higher redshifts (up to $z \approx 0.1$) and can be combined for greater sky coverage.

Anticipating this bounty of new data, we can repeat the above exercise with a mock dataset that reflects these improvements in sky coverage, redshift range, and sample size. For these mocks, we combine the expected redshift and sky distributions of the 50,000 targets selected for the DESI Tully-Fisher peculiar velocity survey \cite{Saudler_2023} with 80,000 galaxy positions and redshifts drawn from the WALLABY reference simulation \cite{Koribalski_2020}\footnote{The WALLABY reference simulation is based on the initial plans for the survey, so its footprint and redshift distribution are now out of date. The full simulation has been downsized appropriately to reflect the current expected Tully-Fisher sample.}.

\begin{figure}[b]
\centering
\includegraphics[width=0.48\textwidth]{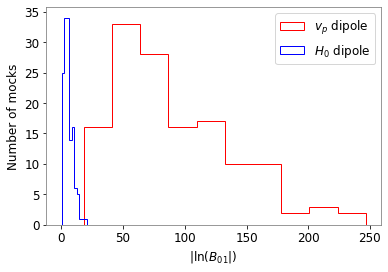}
\includegraphics[width=0.49\textwidth]{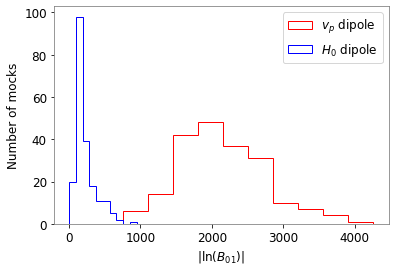}
\caption{Left: Logarithm of the Bayes factor for the correct dipole model relative to the incorrect dipole model in 250 mock CF4 Tully-Fisher datasets. Blue: the true model is an $H_0$ (Tully-Fisher zeropoint) dipole and the median $\ln(B_{01})$ is 2, showing barely any evidence favoring the correct model $M_0$. Red: the true model is a velocity dipole and the median $\ln(B_{01})$ is 79, showing strong evidence favoring the correct model. Right: Forecast of the logarithm of the Bayes factor for the correct dipole model relative to the incorrect dipole model in 250 mock DESI+WALLABY Tully-Fisher datasets. Blue: the intrinsic model is a $H_0$ (Tully-Fisher zeropoint) dipole and the median $\ln(B_{01})$ is 172, showing very strong evidence favoring the correct model $M_0$. Red: the intrinsic model is a velocity dipole and the median $\ln(B_{01})$ is 2090, showing decisive evidence favoring the correct model.}
\label{fig:bayes_future}
\end{figure}

\label{sec:futuremocks}
\begin{figure}[b]
\includegraphics[width=0.4\textwidth]{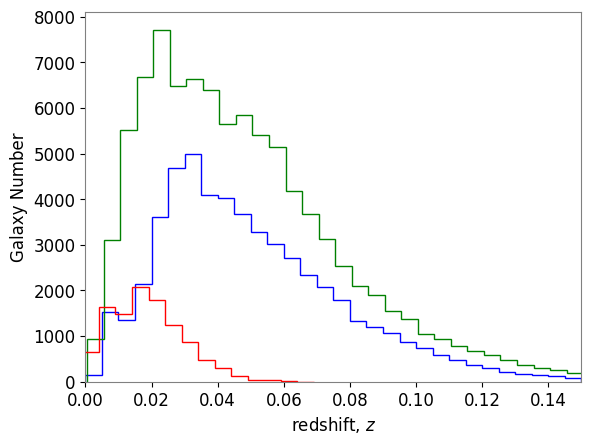} ~
\includegraphics[width=0.58\textwidth]{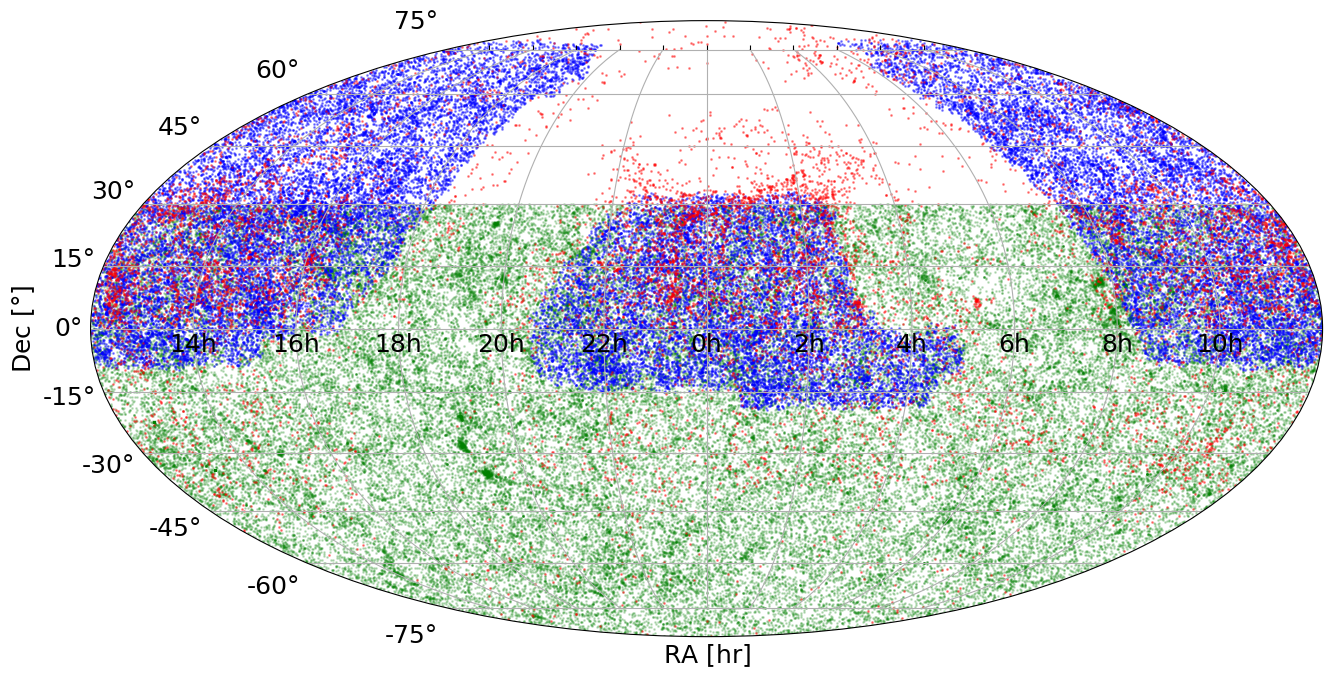}
\caption{The expected redshift and sky distributions for the expected Tully-Fisher data from WALLABY (green) and DESI (blue) compared to existing CF4 data (red).}
\label{fig:desi_wallaby}
\end{figure}

Figure~\ref{fig:desi_wallaby} shows how the redshift and sky distributions of these samples compare to the CF4 Tully-Fisher catalog (WALLABY in green, DESI in blue, CF4 in red). Mock Tully-Fisher data is generated for these mock galaxies using the procedure described in Section~5.2 of \cite{Boubel_2024}. We perform the same tests on these mocks as previously for the CF4 mocks. Again, we assume a 3\% dipole anisotropy in $H_0$ or a velocity dipole with the amplitude of the external bulk flow measured by ~\cite{Carrick_2015}, both in the same direction as the external bulk flow.

As with the mocks in Section~\ref{sec:cf4mocks}, we seek to distinguish a velocity dipole from a $H_0$ dipole using the WALLABY+DESI mocks. For a 3\% $H_0$ dipole amplitude, the measured signal has a significance of 9$\sigma$ when fitting for the $H_0$ dipole. The right panel of Figure~\ref{fig:bayes_future} shows the Bayes factors from the WALLABY+DESI mocks, which very strongly favor the correct model for the anisotropy (whether an $H_0$ dipole or a velocity dipole). Between  CF4  and the WALLABY+DESI mocks, the $\ln{B_{01}}$ values increase by a factor of 25--85. This huge improvement implies future data should much more strongly constrain an $H_0$ dipole, even in the presence of a residual velocity dipole. Even if the strength of the $H_0$ dipole is only 1\%, we find that WALLABY+DESI will be able to detect it with 5.8$\sigma$ significance. 

In the previous section we showed that the complexity introduced by allowing for a quadrupole (in addition to the dipole) was not supported by current data. To gauge whether the future WALLABY+DESI dataset will be large enough to detect a quadrupole anisotropy, we have created a further set of mocks. They are generated in the same way as described above but with an $H_0$ quadrupole anisotropy in addition to the dipole/bulk flow. We find that a quadrupole can be detected with 5$\sigma$ significance if it has an amplitude of at least 1.2\%, in which case the model including the quadrupole is favored over a dipole-only fit with a median Bayes factor $\ln{B_{01}}=814$. This suggests a WALLABY+DESI Tully-Fisher dataset may possibly be able to detect an $H_0$ quadrupole of the form described by \cite{Cowell_2023, Macpherson_2021}, which had an average amplitude of 0.5\% in simulations but a maximum of 5\%.

\section{Conclusions}
\label{sec:conclusions}

We have exploited the ability of the Tully-Fisher relation to test for a direction-dependence of $H_0$ with relatively simple differential zeropoint measurements, due to the fact that only the zeropoint of the relation is affected by changes in $H_0$. Using the best current Tully-Fisher data, the \textit{Cosmicflows-4} catalog, we fit for the Tully-Fisher zeropoint while allowing for variation on the sky in the form of a dipole, a quadrupole (which has some physical motivation \cite{Heinesen_2021, Macpherson_2021, Cowell_2023}), or both (Section~\ref{sec:dipole+quadrupole}).

The best-fit zeropoint dipole, $\widetilde{a_0}$, is ($a_{0x}$, $a_{0y}$, $a_{0z}$)\,=\,($-$0.027\,$\pm$\,0.015, 0.022\,$\pm$\,0.015, 0.048\,$\pm$\,0.014)\,mag with a direction in Galactic coordinates ($\ell$,$b$) = (142\,$\pm$\,30\degree,52\,$\pm$\,10\degree). The dipole amplitude is $|\widetilde{a_0}|$\,=\,0.063\,$\pm$\,0.016\,mag, implying $\Delta H_0$\,=\,2.10\,$\pm$\,0.53\,\kmsMpc\ if $H_0$\,=\,70\kmsMpc, a 3\% variation of $H_0$. If real, this anisotropy would have important implications for the uncertainties in measurements $H_0$ and the Hubble tension. 

The best-fit $H_0$ quadrupole has an amplitude $|\widetilde{a_0}|$\,=\,0.09\,$\pm$\,0.08\,mag, corresponding to $\Delta H_0$\,=\,3.0\,$\pm$\,2.6\,\kmsMpc\ if $H_0$\,=\,70\kmsMpc. This is not a statistically significant result. A Bayes factor of 7.5 when comparing a dipole-only fit to a dipole+quadrupole fit strongly suggests that the improvement to the fit resulting from the quadrupole term is not justified by the additional parameters.

The direction of the dipole does not have any obvious physical significance. It is not aligned with the CMB dipole, so whatever is producing the dipole detected in previous SNIa studies is not present in this Tully-Fisher dataset. It is also not aligned with the Galactic or celestial poles (there is no sign of a north--south miscalibration). However, it is consistent with the direction of minimum anisotropy obtained from galaxy scaling relations \cite{Migkas_2021, Pandya_2024}. 

Assuming no systematic difference in the photometric calibration in different parts of the sky, the other physical effect that may cause anisotropies in $a_0$ is an unaccounted-for velocity dipole. This complicates our analysis because a large residual bulk flow is known to exist \cite{Boubel_2024}. While the effect of a velocity dipole on $a_0$ is different to that of an $H_0$ dipole variation, the two may be difficult to distinguish with insufficient data. In this context, sample size, sky coverage, and redshift range are all important.

From the CF4 data, there is only weak evidence (Bayes factor 0.99) that fitting an $H_0$ dipole in addition to a velocity dipole is favored over a velocity dipole only. This suggests that the existing data is insufficient to constrain an $H_0$ dipole, and that the apparent 3\% variation of $H_0$ on the sky could plausibly be due to a residual bulk flow. With future data, however, our simulations demonstrate that this method can become a powerful tool for the detection of $H_0$ anisotropy.

To generate forecasts for the potential of future datasets, we applied this method to mocks of the expected data from the combined WALLABY and DESI Tully-Fisher surveys. These much larger new surveys significantly tighten the constraints on an $H_0$ dipole relative to a velocity dipole. The anticipated expansion in sample size, redshift range and sky coverage increases the Bayes factor by 25--85 times. This will suffice to detect a 1\% $H_0$ dipole anisotropy with 5.8$\sigma$ significance, a 1.2\% $H_0$ quadrupole anisotropy with 5$\sigma$ significance, and to clearly distinguish an $H_0$ dipole from a velocity dipole of similar amplitude.

\acknowledgments

MMC acknowledges support from a Royal Society Wolfson Visiting Fellowship at the University of Oxford (RSWVF{\textbackslash}R3{\textbackslash}223005). 
KS acknowledges support from the Australian Government through the Australian Research Council Centre of Excellence for Gravitational Wave Discovery (OzGrav) through project number CE230100016.
We acknowledge use of the following analysis packages: Astropy \citep{astropy}, GetDist \citep{getdist}, emcee \citep{emcee}, and Matplotlib \citep{matplotlib}.


\bibliographystyle{JHEP}
\bibliography{biblio.bib}

\end{document}